\documentclass[twocolumn]{aastex6}
\newcommand{\tbf}{}
\pdfoutput=1

\begin{document}

\title{A Substellar Companion to a Hot Star in {\em K2}'s Campaign 0 Field}

\author{S. Dholakia\altaffilmark{1,2,3,9,10}, S. Dholakia\altaffilmark{1,2,3,9,10}, Ann Marie Cody\altaffilmark{2,4}, Steve B. Howell\altaffilmark{2,4,9}, Marshall C. Johnson\altaffilmark{5}, Howard Isaacson\altaffilmark{3}, Mark E. Everett\altaffilmark{6}, David R. Ciardi\altaffilmark{7}, Andrew W. Howard\altaffilmark{7}, Avi Shporer\altaffilmark{7,8}}
\altaffiltext{1}{Wilcox High School, Santa Clara, CA 95051}
\altaffiltext{2}{NASA Ames Research Center, M/S 244-30, Moffett Field, CA 94035, USA}
\altaffiltext{3}{Department of Astronomy, 601 Campbell Hall, University of California, Berkeley, CA 94720, USA}
\altaffiltext{4}{Bay Area Environmental Research Institute, 625 2nd St Ste. 209, Petaluma, CA 94952}
\altaffiltext{5}{Department of Astronomy, The Ohio State University, 140 West 18$^{\mathrm{th}}$ Ave., Columbus, OH 43210}
\altaffiltext{6}{National Optical Astronomy Observatory, 950 N. Cherry Ave., Tucson, AZ 85719, USA}
\altaffiltext{7}{California Institute of Technology, 1200 E California Blvd., Pasadena, CA 91125, USA}
\altaffiltext{8}{Massachusetts Institute of Technology Kavli Institute for Astrophysics and Space Research, 77 Massachusetts Avenue, 37-241, Cambridge, MA 02139, USA}

\begin{abstract}

The {\em K2} mission has enabled searches for transits in crowded stellar environments very different from the original Kepler mission field. We describe here the reduction and analysis of time series data from {\em K2}'s Campaign 0 superstamp, which contains the 150~Myr open cluster M35. We report on the identification of a substellar transiting object orbiting an A star at the periphery of the superstamp. To investigate this transiting source, we performed ground based follow-up observations, including photometry with the Las Cumbres Observatory telescope network and high resolution spectroscopy with Keck/HIRES. We confirm that the host star is a hot, rapidly rotating star, precluding precision radial velocity measurements. We nevertheless present a statistical validation of the planet or brown dwarf candidate using speckle interferometry from the WIYN telescope to rule out false positive stellar eclipsing binary scenarios. Based on parallax and proper motion data from {\em Gaia} Data Release 2 (DR2), we conclude that the star is not likely to be a member of M35, but instead is a background star around 100~pc behind the cluster. We present an updated ephemeris to enable future transit observations. We note that this is a rare system as a hot host star with a substellar companion. It has a high potential for future follow-up, including Doppler tomography and mid-infrared secondary transit observations.

\end{abstract}

\section{Introduction}\label{introduction}

\setcounter{footnote}{8}
\footnotetext{These authors contributed equally to this work}
\footnotetext{Visiting astronomer, Kitt Peak National Observatory, National Optical Astronomy Observatory, which is operated by the Association of Universities for Research in Astronomy (AURA) under a cooperative agreement with the National Science Foundation.}

NASA's $Kepler$ Space Telescope primary mission led to the discovery of 4496 planet candidates over a nearly 4 year span, 3775 of which are now confirmed or validated \footnote{https://exoplanetarchive.ipac.caltech.edu/ as of August 2018}. Following a reaction wheel failure in 2013, the mission was reincarnated as {\em K2}, a series of 70-80 day photometric monitoring campaigns focused on fields in the ecliptic plane \citep{howell2014}. Although the pointing is unstable at the arcsecond level, photometric precision for {\em K2} nevertheless reaches that of the $Kepler$ prime mission down to moderately bright stars ($K_{\rm p}\sim $14--15) \citep{luger2017}. The many observing fields of {\em K2} have enabled not only new exoplanet studies, but science on a diverse collection of astronomical objects, from young stars to pulsators to extragalactic transients. 

In the first full campaign of the {\em K2} Mission, ``Campaign 0,'' the $\sim$150~Myr open cluster M35 (Meibom et al.\ 2008) was targeted in a $\sim$0.6\arcdeg$\times$0.9\arcdeg ``superstamp''-- much larger than the typical $\sim$40$\arcsec$ fields downloaded for individual stars observed by {\em Kepler}. M35 has been the subject of previous searches for exoplanets and variable stars. \citet{nardiello15} presented a long term ground-based study of M35 in preparation for {\em K2}'s Campaign 0, finding 273 new variable stars. Previous ground-based studies had focused on detection of periodic variability for rotation studies \citep{meibom2009}.

In this paper, we present a photometric pipeline for use on the {\em K2} superstamps, and apply it to the M35 dataset from {\em K2} Campaign 0 (Section \ref{k2photometry}). During our analysis of the Campaign 0 data, we found that one of the brighter targets ($V\sim 12.6$) in the M35 field- 2MASS~J06101557+2436535 displayed 0.68\% dips suggestive of substellar transits at period $\sim$7.56 days (Section \ref{discovery}). This candidate was independently reported by \citet{LaCourse15} as part of their Campaign 0 search for eclipsing binary systems. Based on high resolution spectra (Section~\ref{hoststarproperties}), we identify the host as an A type, rapidly rotating star. We use ground-based speckle imaging (Section \ref{speckledata}) to rule out most stellar companion scenarios, and refine the ephemeris with follow-up photometry (Section \ref{LCO}). Finally, we determine a low false-positive probability to confirm the transiting object as a bona fide planet or brown dwarf orbiting 2MASS~J06101557+2436535. 

\section{Photometric Reduction of {\em K2} Postage Stamps}\label{k2photometry}
\subsection{{\em K2} Campaign 0}\label{campaign0}

\begin{figure}[h]
    \centering
    \plotone{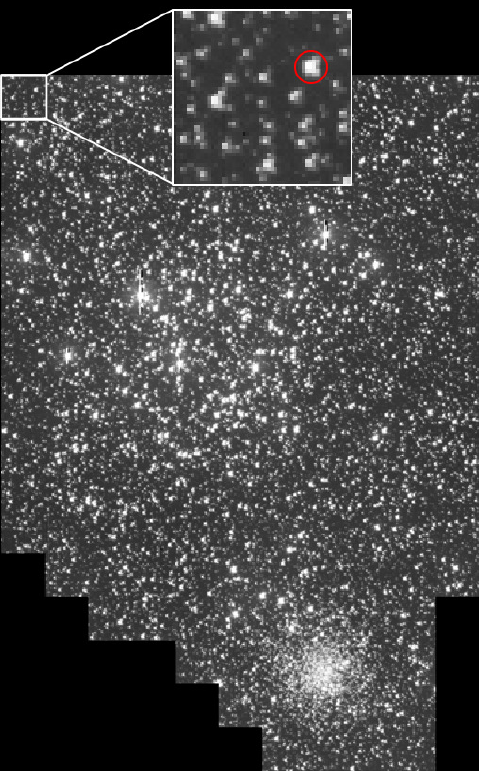}
    \caption{Image showing a {\em K2} Campaign 0 superstamp with an inset depicting the "postage stamp" containing 2MASS~J06101557+2436535. The M35 cluster occupies most of the region, while the older NGC~2158 Cluster is visible toward the bottom. 2MASS~J06101557+2436535 is the star circled in red.} 
    \label{fig:inset}
\end{figure}

Following initial engineering tests, {\em K2}'s Campaign 0 was the first full-length confirmation of the mission's ability to continue acquiring high precision photometric time series with only two working reaction wheels. It lasted from 8 March 2014 to 27 May 2014. Several challenges were encountered during the 79 days of observations. First, data from the initial 40 days were corrupted as the telescope operated in coarse pointing mode, causing excessive photometric scatter. Second, Jupiter passed through the field of view, causing internal reflections and a spike in the background counts lasting approximately 6 hours on 11 May 2014. Third, with the recent failure of a second reaction wheel, the spacecraft slowly rolled along its boresight, requiring corrective thruster firings every 6 hours. Consequently, a gradual dimming and rapid brightening on a six hour period presented itself in the light curves. This is one of the signature systematics of the {\em K2} mission \citep{vanderburg14}. Techniques used to mitigate these effects are detailed in Section \ref{photometryalgorithm}.

Another challenge associated with the {\em K2} mission was the analysis of so-called superstamp data. These images encompass spatially large targets with many stars, such as galaxies and open clusters. For example, {\em K2} Campaign 0 targeted the open clusters M35 and NGC 2158 with a single superstamp, comprised of 154 tiled $50\times 50$ pixel regions; an example superstamp image is shown in Figure~\ref{fig:inset}. Each smaller ``postage stamp'' region encompassed many stars, and the $K2$ mission photometry pipeline did not generate light curves for any of these targets. We therefore created our own set of Python and Pyraf scripts for the reduction of the superstamp data. 

\subsection{Photometric algorithm for light curve generation}\label{photometryalgorithm}

Our {\em K2} superstamp pipeline contains the following steps:

\begin{enumerate}
    \item Identification of stars and generation of aperture masks 
    \item Extraction of light curves
    \item Systematics detrending of light curves
    \item Transit search
\end{enumerate}

\begin{figure}
	\centering
    \epsscale{0.7} 
    \plotone{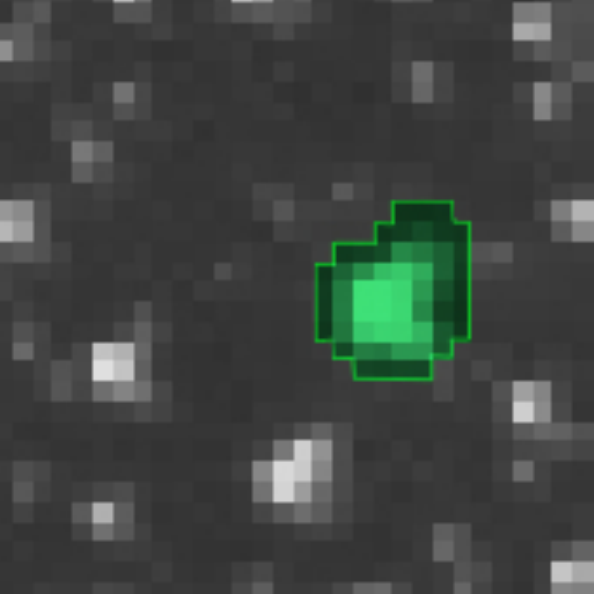}
    \caption{The K2 image region containing 2MASS J06101557+2436535b. The union of the circular and elliptical mask is overlayed in green on the target.}
    \label{fig:maskoverlay}
\end{figure}

First, we obtained 154 FITS format postage stamps of M35 and the adjacent but more distant cluster NGC 2158 from the Mikulski Archive for Space Telescopes (MAST). We generated light curves for 620 stars in total using the procedure described as follows. For all stamps, we located significant local maxima in the images with the tool DAOphot \citep{stetson87} and generated a single aperture mask for each star, an example of which is shown in Fig.~\ref{fig:maskoverlay}, within which we summed the flux to generate photometric measurements. The masks were created from the data in one representative image, taken from the approximate center of each postage stamp's time series. They were then applied to all images in the series. The size and center of each mask was also determined with the help of the IRAF tool DAOphot \citep{stetson87}. We measured a centroid with DAOphot, and subsequently fit the point spread function (PSF) of the star using the IRAF task PSFmeasure. Then, the FWHM value output by PSFmeasure was used to compute the radius of a circular mask using a scaling relationship developed through experimentation. Optical aberration at the edge of the field, where the M35 superstamp was located, caused elongation of the PSFs. To account for this asymmetric shape, we experimented with the addition of a slightly larger elliptical mask to the circular mask determined above. We found that the union of the circular mask based on the PSF size and an elliptical mask with angle and elongation determined by eye encompassed stars' flux well while excluding background and other nearby stars in more crowded fields typical of the M35 superstamp. We used this mask shape, scaled for star brightness, on every star 7 sigma above the background threshold, in the C0 M35 field. Crowding was far more severe in the region surrounding the cluster NGC 2158; we do not attempt to search for planets in the most crowded parts of this region.

The PyKE \citep{still12} tool kepextract was used to create light curves by extracting the flux in every pixel within the mask over the length of the timeseries and applying a sky subtraction. We then applied several refinement techniques to the light curves to remove systematics and artifacts. Initial detrending of light curves was performed using the PyKE tool kepflatten. Data flagged as poor quality due to thruster firings were removed. The PyKE tool \textit{kepsff}, based on the self flat fielding algorithm of \citet{vanderburg14} was used to mitigate the dominant {\em K2} bore-sight roll systematic effect, which appears as a sawtooth pattern in the light curves. The PyKE tool \textit{kepbls} based on \citet{kovacs2002} was then used to generate box least-squares fits to search the sample space for period, transit duration, and transit depth, returning a box least squares (BLS) periodogram. The peak normalized signal residual was compared to the median absolute deviation to find a signal detection efficiency (SDE) as described in \citet{kovacs2002}. An SDE above 6 was used as an initial flag for potential sources of astrophysical variability similar to transits.

\subsection{Identification of a transiting substellar object}\label{discovery}

\begin{figure*}[h]
    \centering
    \plottwo{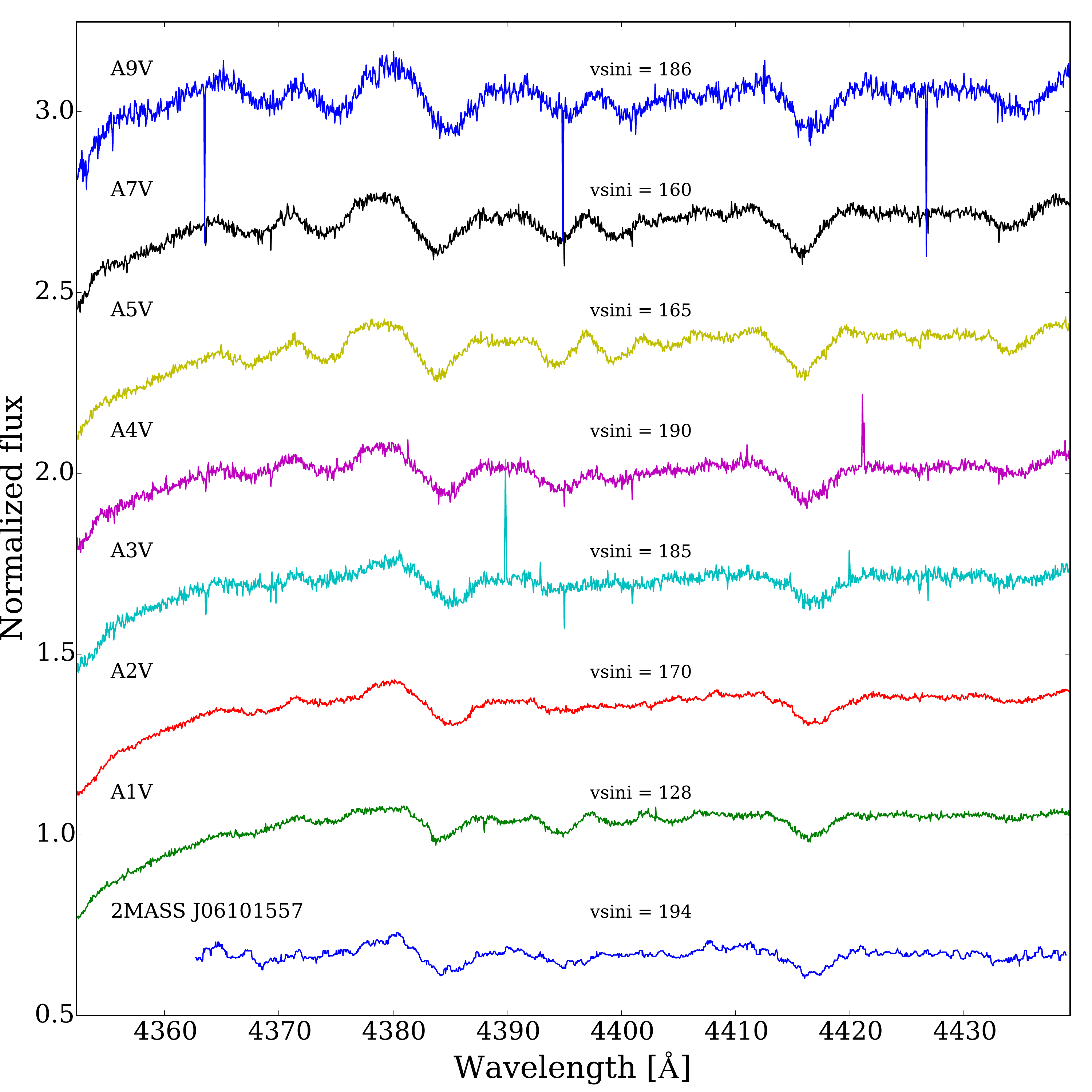}{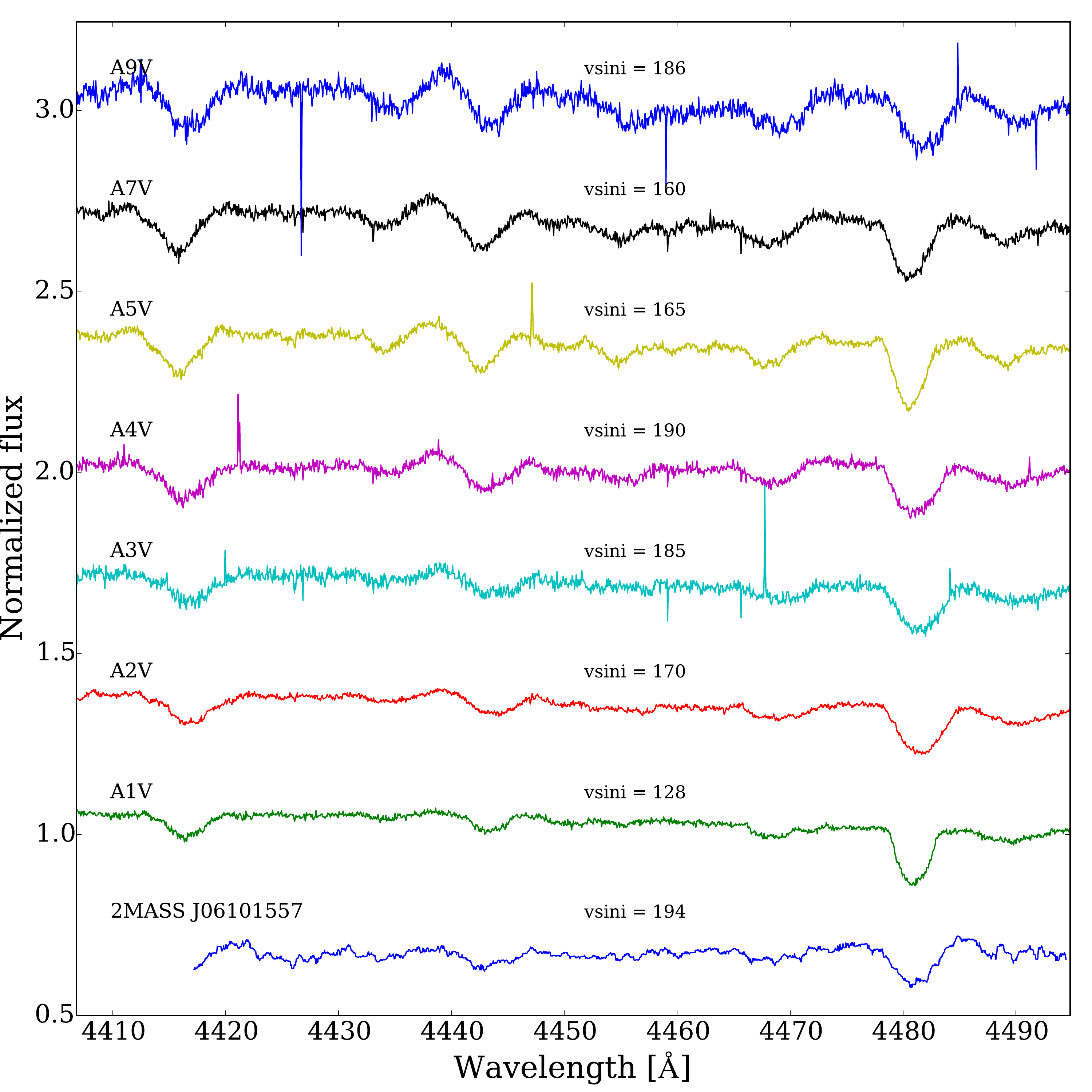}
    \plottwo{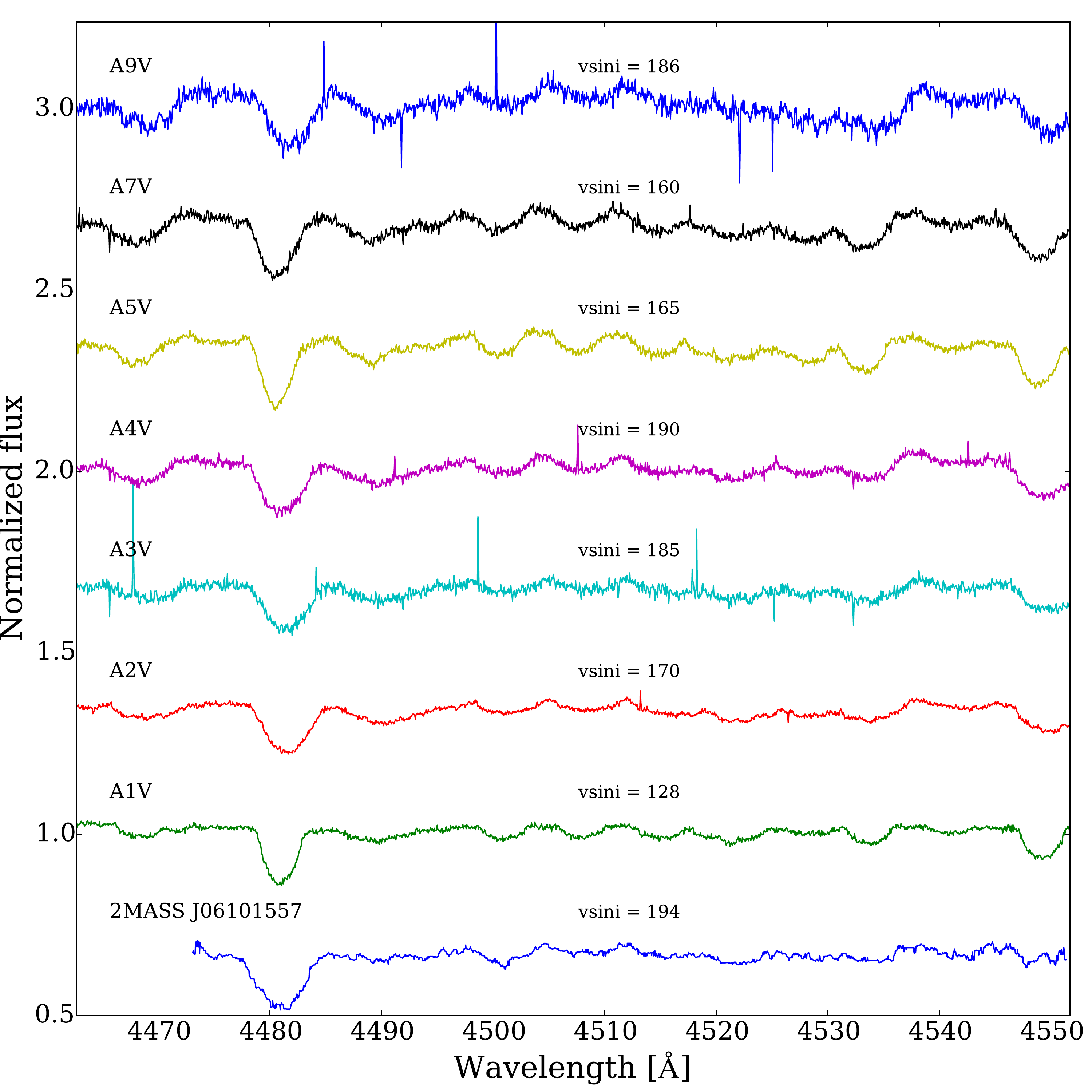}{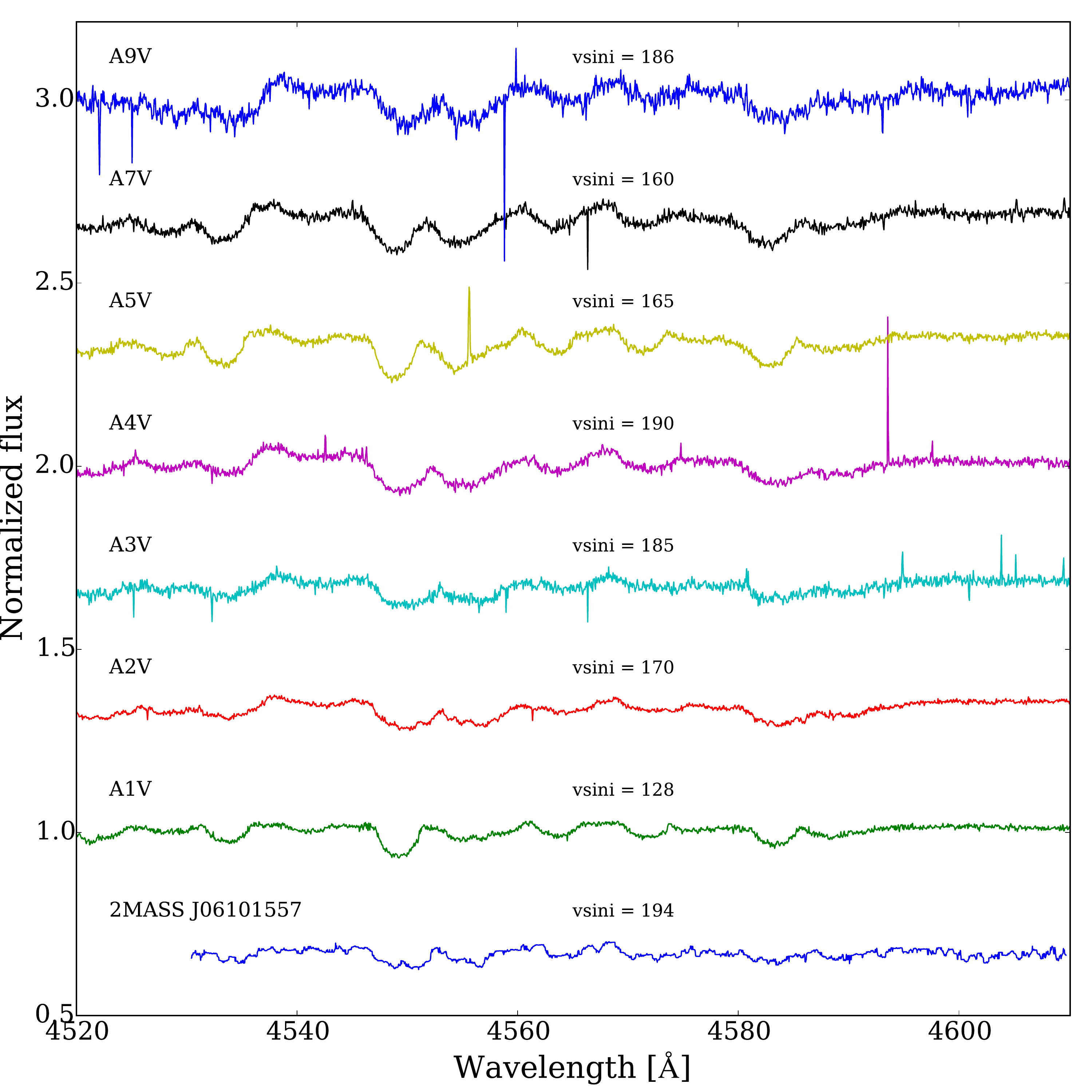}
    \caption{We show several orders of the Keck spectra for 2MASS~J06101557+2436535 (bottom; blue), as compared with a sequence standard star spectra from the ELODIE library. The sequence runs from A1 to A9 and consists of stars with large $v\sin i$. We have used it to determine that the features seen in our object spectrum are the closest match to an A2 or A3 spectral type.} 
    \label{fig:keckspectra}
\end{figure*}

\begin{figure}[h]
    \centering
    \plotone{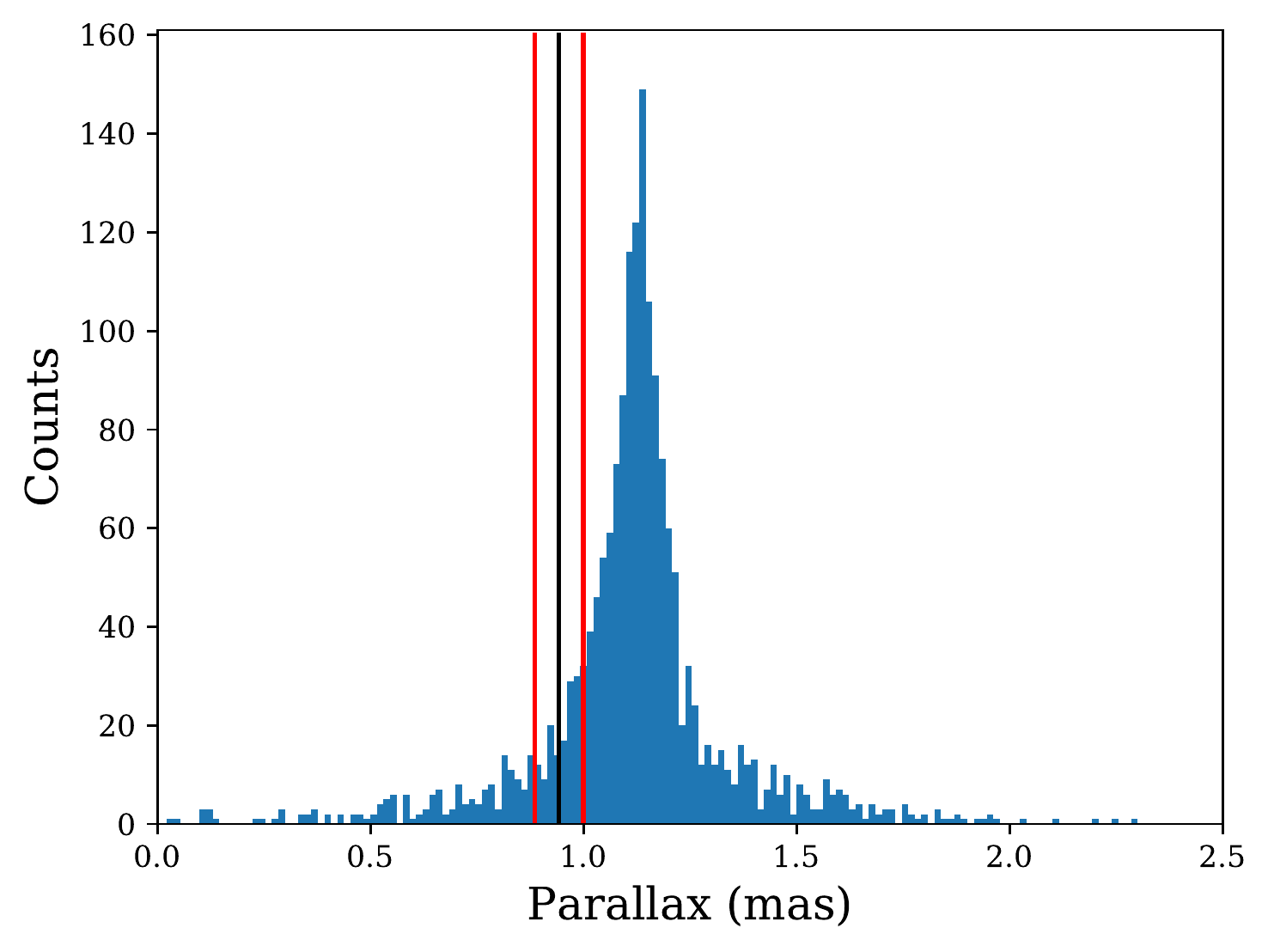}
    \caption{Histogram showing distribution of parallaxes from stars included as M35 members in \citet{babusiaux2018}. The black line marks the 2MASS~J06101557+2436535's parallax in {\em Gaia}, which lies beyond the cluster mean distance and therefore presents a low probability of membership. The red lines represent the errors.} 
    \label{fig:gaiaparallax}
\end{figure}

\begin{figure}[h]
    \centering
    \plotone{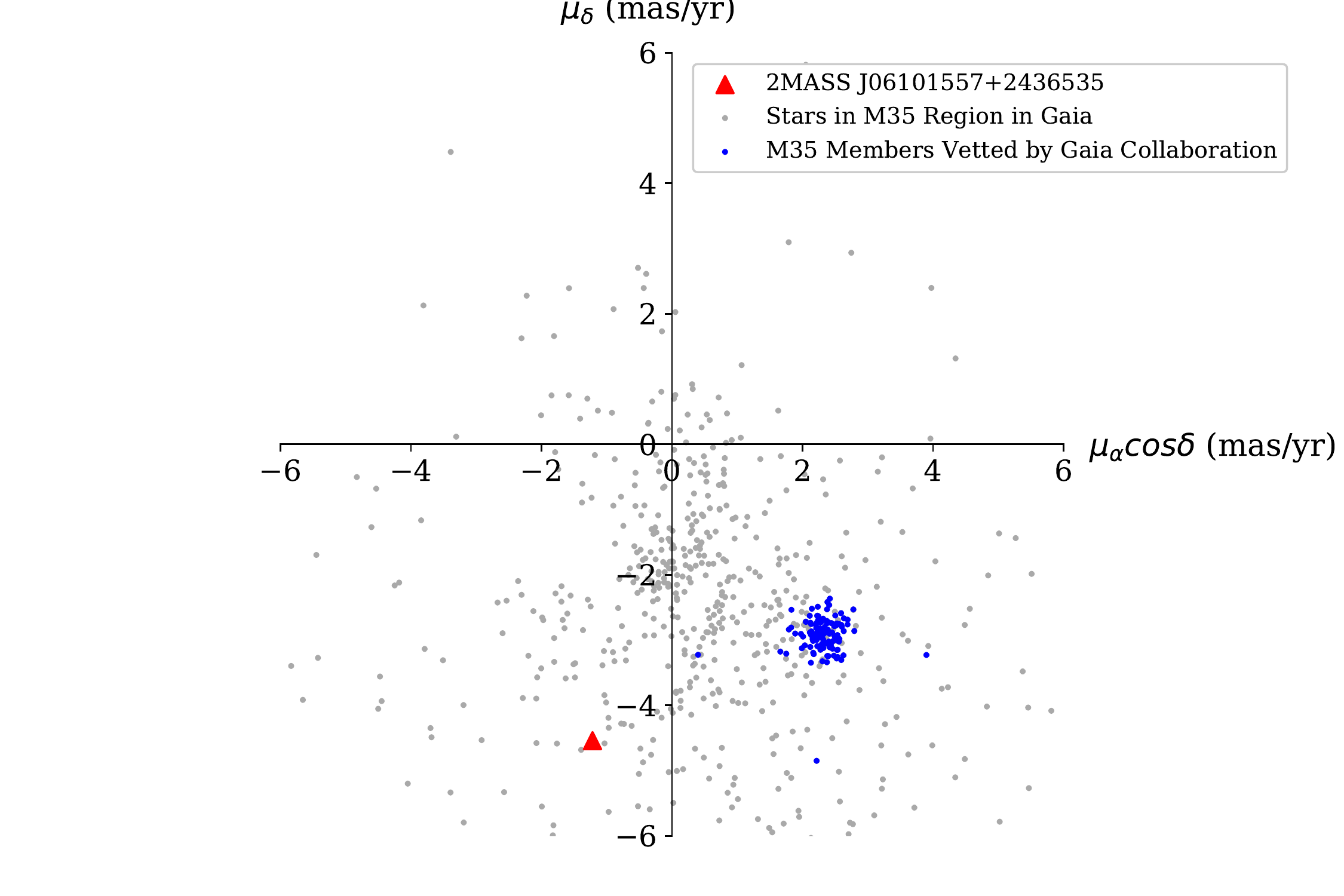}
    \caption{Proper motions in {\em Gaia} of M35 members and nonmembers showing 2MASS~J06101557+2436535. Members in blue are stars vetted by \citet{babusiaux2018} as members of M35. Nonmembers in grey are stars in a 1 degree radius around the M35 center that do not appear in the aforementioned list. All stars are within 0.5 of the 2MASS~J06101557+2436535 Gaia G magnitude. The red triangle marks the position of 2MASS~J06101557+2436535. Error bars for its proper motion are insignificant on this scale.}
    \label{fig:gaiapm}
\end{figure}

\begin{figure}[h]
    \centering
    \includegraphics[width=0.5\textwidth]{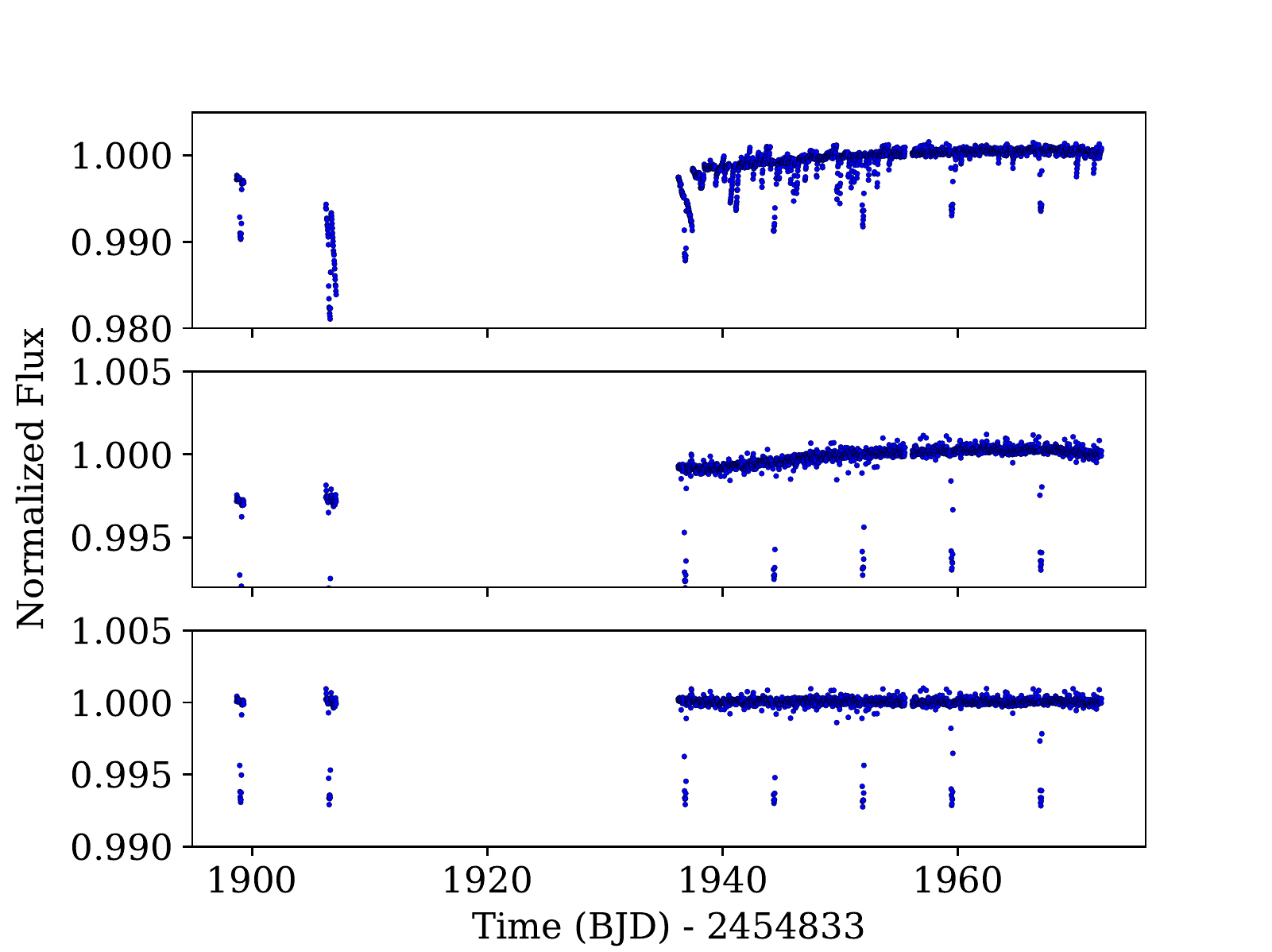}
    \caption{Raw lightcurve, pointing detrended lightcurve and fully detrended lightcurve (i.e., all out-of-transit variability removed) for 2MASS J06101557+2436535b. Data is missing in the first half of the light curve because {\em Kepler} operated in coarse point mode, and precision was greatly reduced and many images were corrupted. Note the light curves are plotted on different y axes.} 
    \label{fig:fullkeplerlc}
\end{figure}

\begin{figure}[h]
    \centering
    \plotone{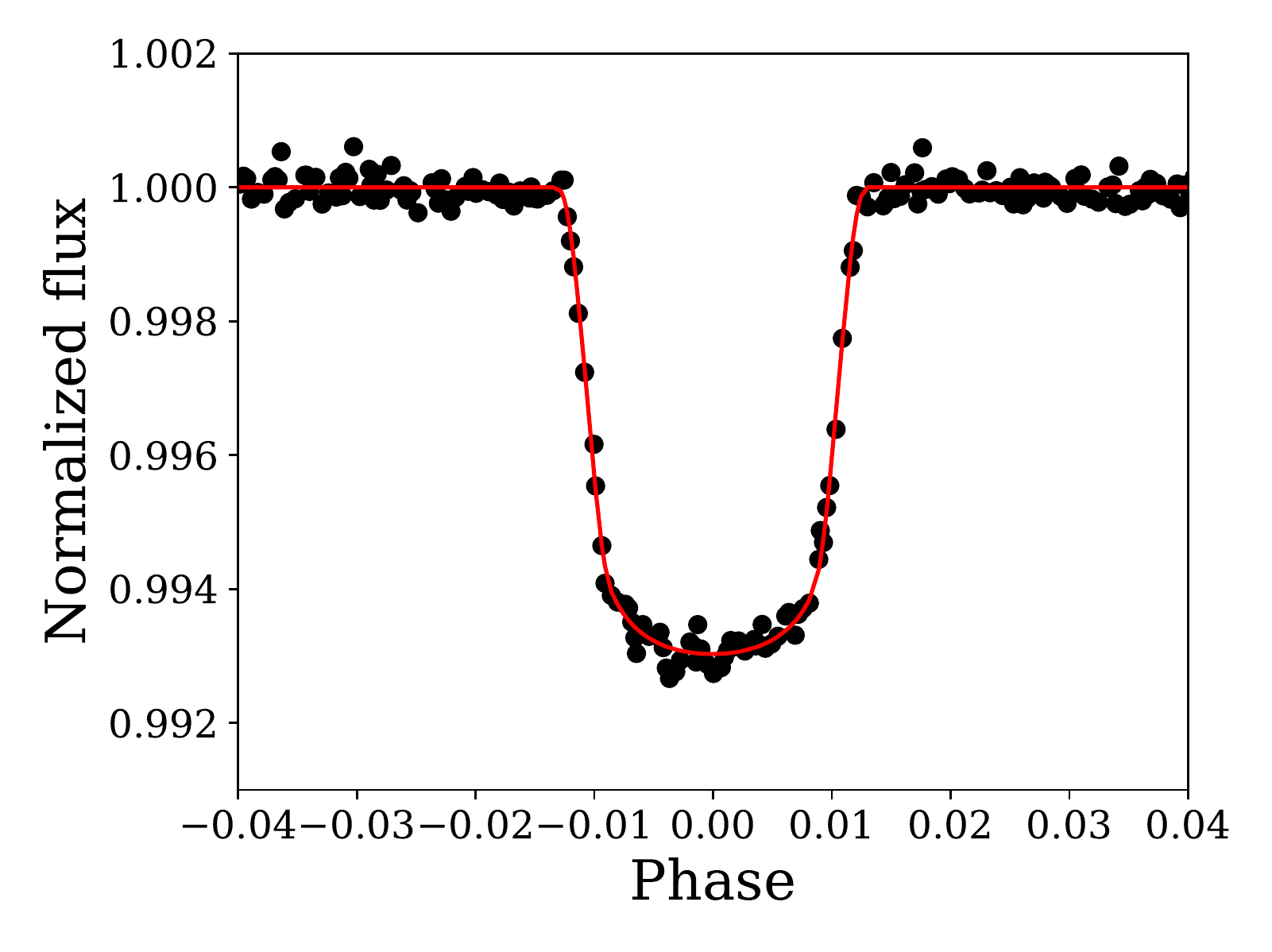}
    \caption{Phase-folded light curve of the transit of 2MASS J06101557+2436535b, with a period of ~7.56 days. The black points are the {\em K2} photometry data. The best fitting transit model from EXOFAST is shown in red.} 
    \label{fig:keplerlc}
\end{figure}

In our analysis of the {\em K2} Campaign 0 superstamp, we found that the light curve of the relatively bright ($V=12.6$) star 2MASS~J06101557+2436535 displayed five 0.68\% dips indicative of a possible transiting substellar object. This star has received relatively little attention in the literature, apart from inclusion in all-sky photometric surveys (see Table~1) and an independent report of a possible planetary companion by \citet{LaCourse15}. \tbf{Photometric searches of the M35 superstamp by \citet{libralato2016} did not report this substellar companion as it was outside their analyzed field of view. \citet{soaresfurtado2017} extract light curves for stars including 2MASS~J06101557+2436535 but do not analyze or discuss the lightcurves individually.} We also recovered several eclipsing binaries and three other potential substellar candidates. However, because of their v-shaped transits, the latter were likely to be false positives and we did not follow them up. We describe characterizations of the substellar companion in Sec.~\ref{companioncharacterization}.

\begin{figure}
    \epsscale{1.3}
    \plotone{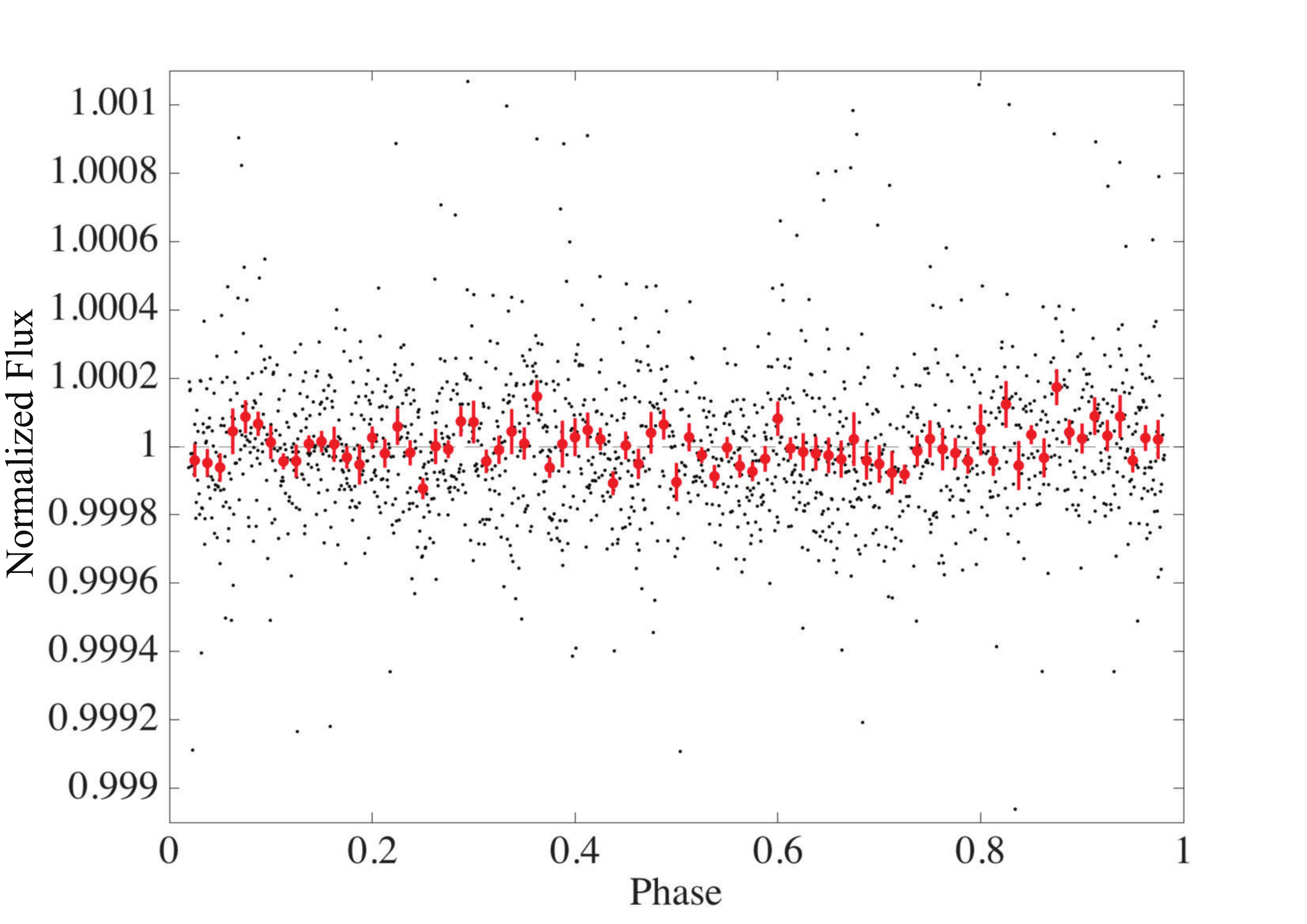}
    \caption{Phase folded {\em K2} light curve (black points) overplotted by the binned light curve (red), where the bin width was taken to be half the transit duration. A dashed horizontal line is plotted at unity for reference. In-transit data is excluded from this plot. The scatter in the un-binned light curve is 0.022\% and in the binned light curve 0.006\%.}
    \label{fig:phase}
\end{figure}

\begin{figure}
    \epsscale{1.2}
    \plotone{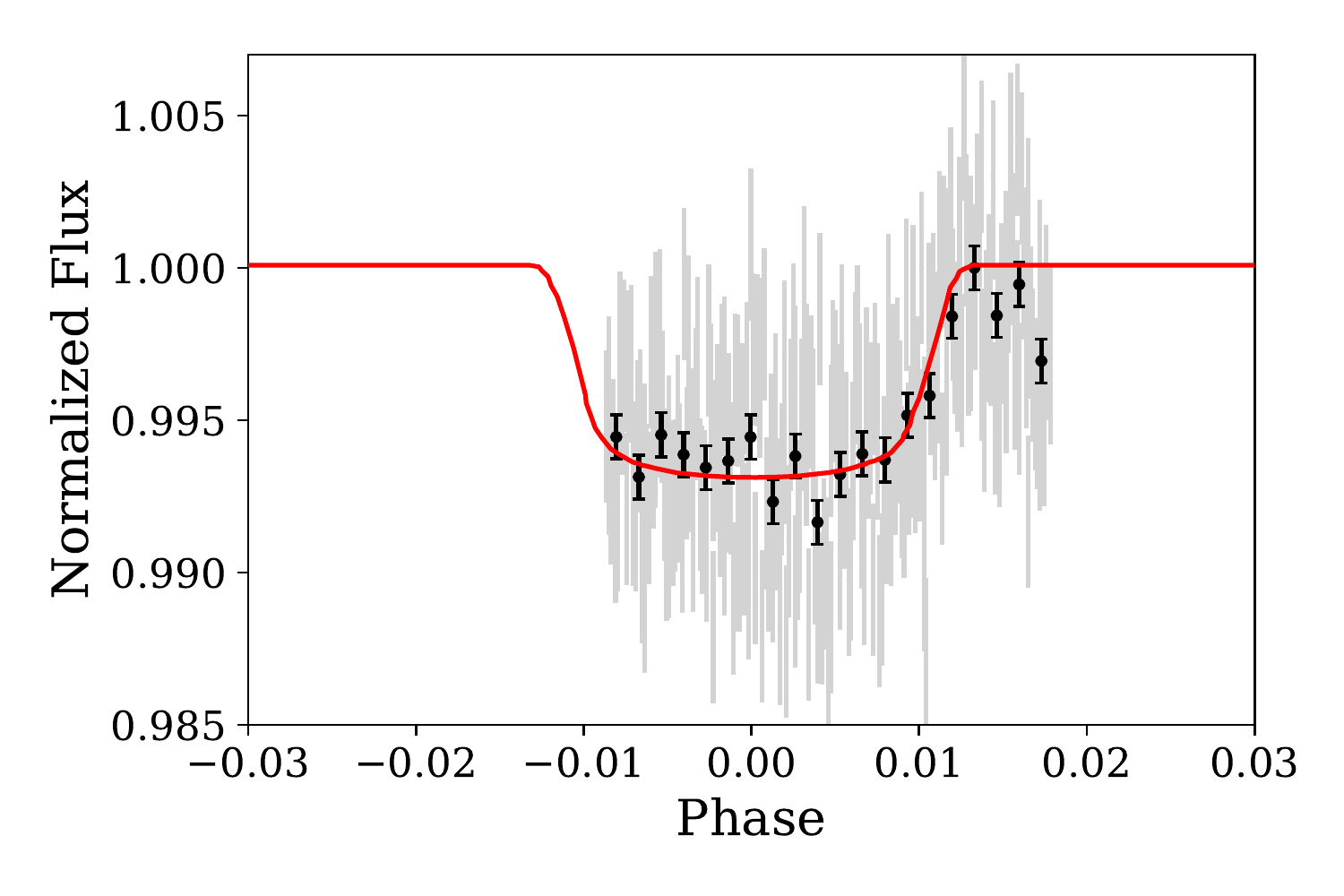}
    \caption{Partial transit observed with the LCO 0.4m telescope on Haleakala approximately 2 years and 8 months after the original {\em K2} observations. The unbinned data are shown in grey, the black points are binned by 12, and the red line is the same EXOFAST transit model fit described towards the end of Section \ref{LCO}.}
    \label{fig:lcolc}
\end{figure}

\section{Host star properties}\label{hoststarproperties}

2MASS~J06101557+2436535 is a moderately bright ($V=12.6$) yet little studied star on the periphery of the M35 cluster field. We gathered photometry from the Two Micron All Sky Survey (2MASS) and the USNO CCD Astrograph Catalog (UCAC), versions 4 and 5, as shown in Table~1. We also adopted proper motions from {\em Gaia}'s second data release (``{\em Gaia} DR2''), which we use for membership assessment in Section \ref{membership}. \tbf{\citet{LaCourse15} estimated a spectral type of A2IV/V, although this was from a low-resolution spectrum.} To gain further insight into the host star's properties, and to estimate the star's projected rotational velocity (Section \ref{vsini}), we conducted high resolution spectroscopy using the Keck Observatory High Resolution Echelle Spectrometer (HIRES).

\subsection{Spectral type} \label{spectraltype}
Several high resolution spectra were collected with the Keck HIRES instrument, using the standard California Planet Search setup \citep{howard2010}. Observations were taken on 21 August 2015, 10 October 2016, 17 October 2016, and 11 November 2016. During each of the observation dates, seeing was approximately 1 arcsecond. The observation taken on 21 August 2015, with an exposure time of 180s did not have the iodine cell in the  light path, allowing for analysis across the entire spectral format, spanning 3700 to 8000 Angstroms. The signal to noise ratio of each spectrum was 45 at 5500~\AA. Using the 0.87$\arcsec\times 14.0\arcsec$ decker, the resulting spectral resolution is $\sim$60,000. Sky subtraction was performed removing spectral emission lines due to Earth's atmosphere. 

The HIRES spectra exhibited strong Doppler broadening and few absorption lines, indicating a rapidly rotating hot star. To estimate a spectral type, we compared our spectrum with those of template A and F stars in the ELODIE library \citep{prugniel2007}; several spectral orders are shown in Fig.\ \ref{fig:keckspectra}\tbf{.} Since 2MASS~J06101557+2436535 is rapidly rotating (see Section~\ref{vsini}), we only compared its spectrum with those of stars with $v\sin i >150$~km/s. Our target exhibits rotationally broadened lines consistent with a spectral type of A2--A3. Accordingly, we adopt an effective temperature of 8500$\pm 500$~K, based on the median and standard deviation of the temperatures reported for our A2 and A3 standard stars from the ELODIE library.

\subsection{Rotational velocity} \label{vsini}

We measured 2MASS~J06101557+2436535's rotational broadening using the \texttt{misttborn} code\footnote{\url{https://github.com/captain-exoplanet/misttborn}} to fit a stellar line model, generated as described in \citet{johnson2014}, to a single line from the first of our HIRES spectra. This model assumes a Gaussian line profile from each stellar surface element (incorporating thermal and microturbulent broadening, and an approximation of macroturbulent broadening), and Doppler shifts the line profiles from each surface element assuming solid body rotation to produce a rotationally-broadened model line profile. We then fit this model to the data using \texttt{emcee} \citep{emcee2013}, setting Gaussian priors on two quadratic limb darkening coefficients and using the triangular sampling method of \citet{kipping2013} for these parameters. We ran 100 walkers for 150,000 steps each. This produced a projected stellar rotational velocity of $v\sin i=193.7_{-6.4}^{+6.2}$~km/s and a non-rotating Gaussian line width of $47.9_{-14.6}^{+14.4}$~km/s.

\subsection{Radius} \label{radmass}
Using the parallax provided by {\em Gaia} DR2, we can derive its radius based on effective temperature and luminosity. We computed the star's bolometric luminosity from the $V$-band magnitude, and an age-insensitive $V$-band bolometric correction of -0.08$\pm$0.02 \citep{pecaut2013}. We adopt a distance 1061$^{+69}_{-61}$~pc from the {\em Gaia} DR2 parallax of 0.9424$\pm 0.0573$~pc. We estimate the reddening by assuming it is close to that of M35 \citep[$E(B-V)=0.20$ or $A_V=0.62$;][]{kalirai2003}. This is supported by the similar value reported by {\em Gaia} for the $G$-band reddening, which is 0.69. The resulting absolute $V$ magnitude is $M_V=1.82\pm 0.16$. This absolute magnitude results in a luminosity of $L_*=15.84^{+2.6}_{-2.2}$~$L_\odot$. 

{\em Gaia} also lists an effective temperature of $7085.00^{+113}_{-244}$ K. However, \citet{andrae2018} notes that the bluer stars have systematically underestimated T\textsubscript{eff}, and as such we adopt the T\textsubscript{eff} derived from the spectrum. This is supported by our comparison of the star's spectrum with the ELODIE standards in Fig.\ \ref{fig:keckspectra}; it is clear that our target is {\em not} a late A or early F star.

With the estimated spectral type of A2--A3 and corresponding effective temperatures of 8500$^{+500}_{-500}$~K, we obtain a stellar radius of $R_*=1.85^{+0.4}_{-0.3}$~$R_\odot$. This value is in line with the radii predicted by the models of \citet{siess2000} for a star with spectral type A2--A3.

As per \citet{seager2003}, the mean density of the host star can be determined with solely a transit measurement, although this measurement is degenerate with the impact parameter and eccentricity of the transiting object. We obtained a stellar mean density of $<\rho> = 0.31^{+0.17}_{-0.13}$~g/cm$^3$ from our transit fit, described in further detail in Section \ref{transitfit}.

\begin{deluxetable}{lll}
\tablecolumns{3}
\tablewidth{0pt}
\tablecaption{2MASS J06101557+2436535 Properties}
\tablehead{
\colhead{Parameter} & \colhead{Value} & \colhead{Source}
}
\startdata
  $\alpha$ R.A. (hh:mm:ss) & 06:10:15.57 & Gaia \\
  $\delta$ Dec. (dd:mm:ss) & 24:36:53.38 & Gaia \\
  $\mu_{\alpha}cos(\delta)$ (mas/yr) & -1.218	 $\pm$ 0.085 & Gaia \\
  $\mu_{\delta}$ (mas/yr) & -4.542 $\pm$ 0.072 & Gaia \\
  $B$ (mag) & 13.02 $\pm$ 0.03 & UCAC4 \\
  $V$ (mag) & 12.57 $\pm$ 0.03 & UCAC4 \\
  $G$ (mag) & 12.55 $\pm$ 0.00 & {\em Gaia} \\
  $J$ (mag) & 11.77 $\pm$ 0.02 & 2MASS \\
  $H$ (mag) & 11.68 $\pm$ 0.02 & 2MASS \\ 
  $K$ (mag) & 11.56 $\pm$ 0.02 & 2MASS \\ 
  $v\sin i$ (km/s) & $193.7_{-6.4}^{+6.2}$ & This work\\
  Spectral Type  & A2-A3 &  This work\\
  Luminosity ($L_\odot$) & 15.84$^{+2.6}_{-2.2}$ &  This work\\
  Mean stellar density (cgs) & $0.31^{+0.17}_{-0.13}$ & This work \\
  Radius ($R_\odot$) & $1.85^{+0.4}_{-0.3}$ & This work\\
 \enddata
\tablecomments{\label{tab:stellarparameters} Host star properties from the literature and derived in this work.}
\end{deluxetable}

\section{Cluster membership} \label{membership}
Two prior membership surveys of M35 contained this star, \citet{cudworth71} and \citet{bouy15},  both of which listed a membership probability of 0. However, the \citet{cudworth71} work was based entirely on proper motions from two photographic plates with potentially large astrometric uncertainties. Our investigation showed that in general, their proper motions for M35 stars in the x direction of the plate pairs vary greatly from the values reported in UCAC5, especially for 2MASS~J06101557+2436535. Further, we discovered that \citet{bouy15}'s adopted Sloan Digital Sky Survey photometry for 2MASS~J06101557+2436535 was saturated. As a result, we were compelled to assess this star's M35 membership status using the high precision astrometric measurements from {\em Gaia} DR2 \citep{gaiacollab2016}, \citep{gaiacollab2018}. This catalog provides precise measurements for both parallax and proper motion, allowing an analysis of cluster membership both using distance to the cluster center and proper motions relative to other members.

2MASS~J06101557+2436535 is listed in {\em Gaia} DR2 as G2~3426303357158956544. The parallax value is listed as $0.9424\pm0.0573$~mas, corresponding to a distance of 1061$^{+69}_{-61}$~pc. This is beyond the M35 cluster center distance of 885~pc, according to the {\em Gaia} DR2 mean cluster parallax of $1.1301 \pm0.0013$~mas \citet{babusiaux2018}, \tbf{as shown in Fig.~\ref{fig:gaiaparallax}}. We also note that these Gaia parallaxes and the distances derived from them may be affected by position, magnitude, and color dependant offsets.

{\em Gaia} DR2 also reports astrometric data which yield measurements of a source's proper motion. We compared the proper motion of 2MASS~J06101557+2436535 to the cluster’s proper motion. The star's proper motion of $-1.218$~mas/yr in R.A. and $-4.542$~mas/yr in Dec. are far from the well-defined cluster mean at $2.2784\pm{0.0052}$~mas/yr in R.A. and $-2.9336\pm{0.0050}$~mas/yr in Dec. These values for the cluster mean are presented in \citet{babusiaux2018}. \tbf{The clustering of proper motions for M35 members are shown in Fig~\ref{fig:gaiapm}.}

Analyses of cluster membership based on parallaxes and proper motions has been presented in \citet{babusiaux2018} (see their Table A1b for a list all members of M35 based on {\em Gaia} parallaxes and proper motion). We find that 2MASS~J06101557+2436535 is not in the list of members, likely owing to its distance from the cluster center and different proper motions, and we therefore do not consider it a member of M35.


\section{Characterization of Substellar Companion} \label{companioncharacterization}

We undertake a detailed characterization of the substellar companion \tbf{using} all available sources of data in order to determine the planetary characteristics. As described in Section \ref{spectraltype}, 2MASS~J06101557+2436535 is rapidly rotating, presenting Doppler broadened lines that preclude a constraint of the radial velocity and mass of the substellar companion. We pursue other methods in order to constrain the mass to the extent possible through other means and validate that the companion is indeed substellar and not a false positive.

\subsection{Transit parameters} \label{transitfit}

To perform a detailed assessment of the parameters of 2MASS~J06101557+2436535 and the transiting object, we returned to our raw {\em K2} light curve and removed systematic trends using the {\em k2sc} detrending algorithm provided by \citet{aigrain2016}. {\em k2sc} uses both pixel-level decorrelation to remove spacecraft associated systematics and Gaussian processes to reduce astrophysical variability and also allows transits to be masked in the detrending to avoid reducing the transit depth during detrending. \tbf{The {\em k2sc} detrending is shown in Fig~\ref{fig:fullkeplerlc}.} We fit the resulting detrended light curve with a Markov Chain Monte Carlo (MCMC) model using the EXOFAST \citep{eastman2013} package as provided by the NASA Exoplanet Archive, \tbf{which uses the \citet{mandel2002} model}. The best fit model is shown in Figure \ref{fig:keplerlc}.  Ultimately we discarded most of the out of transit data during the coarse point part of Campaign 0. However, two transits in the initial part of the light curve for 2MASS~J06101557+2436535 retained enough precision to be included in the fits. In total, seven transits were fit. The best fit included a period of 7.556 days and a depth of 0.6839\%, which corresponds to an $R_p/R_*$ of 0.08270. This corresponds to a planetary radius of 1.5$\pm0.3$~$R_J$. The full list of parameters obtained from the transit fit is presented in Table~2.

\subsection{Photometric mass limits}\label{photmasslimit}

The transit depth and lack of a secondary eclipse at any phase can be used to constrain the possibility of a stellar companion to the A star host. Assuming that the host star is not in M35 (see Sec. \ref{membership}, a typical A3 star radius of 1.85~$R_\odot$ (see Section~\ref{radmass})) yields an eclipse depth of $\sim$0.65\% for a 0.1~$M_\odot$ star-- just below what is observed. These estimates rely on the assumption that we are not viewing a grazing eclipse. Such a scenario is unlikely, as it would produce more of a V-shaped dip than observed; we statistically quantify this situation in Section~\ref{validation}. We conclude that the transiting companion almost certainly must be substellar.

To search for a secondary eclipse, we examined out-of-transit data in the phase folded light curve as shown in Figure~\ref{fig:phase}. The scatter in this data is 0.022\%. The light curve was binned using bin width of half the transit duration and no obvious secondary eclipse is seen in the binned light curve. We estimate an upper limit on the secondary eclipse depth of 0.020\%, which is above the largest absolute deviation of any of the bins from unity, and close to three times the scatter in the binned light curve, of 0.006\%. Given the properties of the primary star (see Section~\ref{hoststarproperties}) an eclipse depth of 0.020\% is expected from a mid M-dwarf star with an luminosity of $0.003L_\odot$. \tbf{Given the luminosity of our star from Table. 1 and the definition of the eclipse depth which is $\frac{L_{companion}}{L_{star}+L_{companion}}$, the companion must be fainter than 0.02\%, which can be converted to mass using the \citet{siess2000} model.}
This results in a companion mass of about $\sim$0.2~$M_\odot$. It should be noted, however, that if a significant nonzero eccentricity exists in the system, it is possible for transits to occur but not secondary eclipses.

Another way to constrain the companion mass is by looking for photometric orbital modulations along the orbital phase induced by the orbital motion of the two objects around the common center of mass \citep[e.g.][]{zucker07, faigler11, shporer17}. Since the phase curve does not show any variability down to the noise level (see Figure~\ref{fig:phase}), we applied injection and recovery of an orbital signal to place an upper limit on the transiting object mass, while assuming a circular orbit. For a given trial companion mass we injected the corresponding photometric orbital signal into the light curve, using the same period and phase as the transit. We then tested whether the signal is recovered by generating the Lomb-Scargle periodogram \citep{scargle82} and requiring for detection that the strongest periodogram peak be at the orbital period (or its first harmonic) and be at least twice as high as the 2nd strongest periodogram peak (which is not a harmonic of the orbital period). Assuming a host star with an effective temperature of 8500~K and radius and mass of 1.85~$M_\odot$, the companion mass upper limit is 96~$M_{\rm J}$, or 0.092~$M_\odot$.  Therefore, this analysis places an upper limit on the mass of the companion, suggesting it is close to the hydrogen burning mass threshold, at the bottom of the main sequence, or substellar.

\subsection{Ground-based transit photometry and improvement of ephemeris} \label{LCO}

In order to confirm the long-term stability of the light curve behavior and improve the transit ephemeris, we carried out four hours of ground based photometric monitoring in the SDSS r' band using the Las Cumbres Observatory (LCO) telescope network \citep{brown2013} on 5 Feb 2017 (UT). A transit was recovered in data taken with the LCO 0.4m telescope and SBIG STX6303 CCD camera on Haleakala, Hawaii. This transit occurred during twilight, so photometry for only the latter half of the event was collected.  We took 241 sequential 60s exposures with an SDSS r' filter. These were reduced using AstroImageJ \citep{astroimagej}, an interactive astronomical image processing and photometry software built on the Java-based software ImageJ. AstroImageJ allows users to place circular annulus apertures onto a reference image, then aligns and performs aperture photometry, generating light curves for a target star. We used a 9, 20 and 32 pixel radius respectively for the aperture, inner annulus and outer annulus. Because ground based photometry can be greatly affected by airmass, telescope systematics and other issues, careful selection of comparison stars is crucial for optimal differential photometry precision. Eight comparison stars ranging from 11th magnitude to 14th magnitude were used in the reduction of the image. 

The final light curve was produced by binning sets of 12 points, as show in Figure~\ref{fig:lcolc}. Error bars were determined from both photon and background shot noise. The LCO partial transit was added to the {\em K2} data to create a longer baseline time series. To reduce the uncertainty on the system's period, we input the light curve to EXOFAST \citep{eastman2013} using priors from the EXOFAST model found in Section \ref{discovery}. We retained most of the original EXOFAST fit values from the fit to {\em K2} data alone (Table 2), because the LCO data included only one partial transit and had a low signal-to-noise ratio relative to the {\em K2} observations. However, it did help in refining the ephemeris and reducing the error bars on the period. The period value obtained from the fit of both the {\em K2} transits and LCO partial transit was $ 7.555907^{+0.000029}_{-0.000030}$~days. Figure~\ref{fig:lcolc} shows the EXOFAST transit model overplotted on this ground based photometry; it corroborates the transit depth and improves the period found from the {\em K2} observations.

\begin{figure}
	\centering
    \epsscale{1}
    \plotone{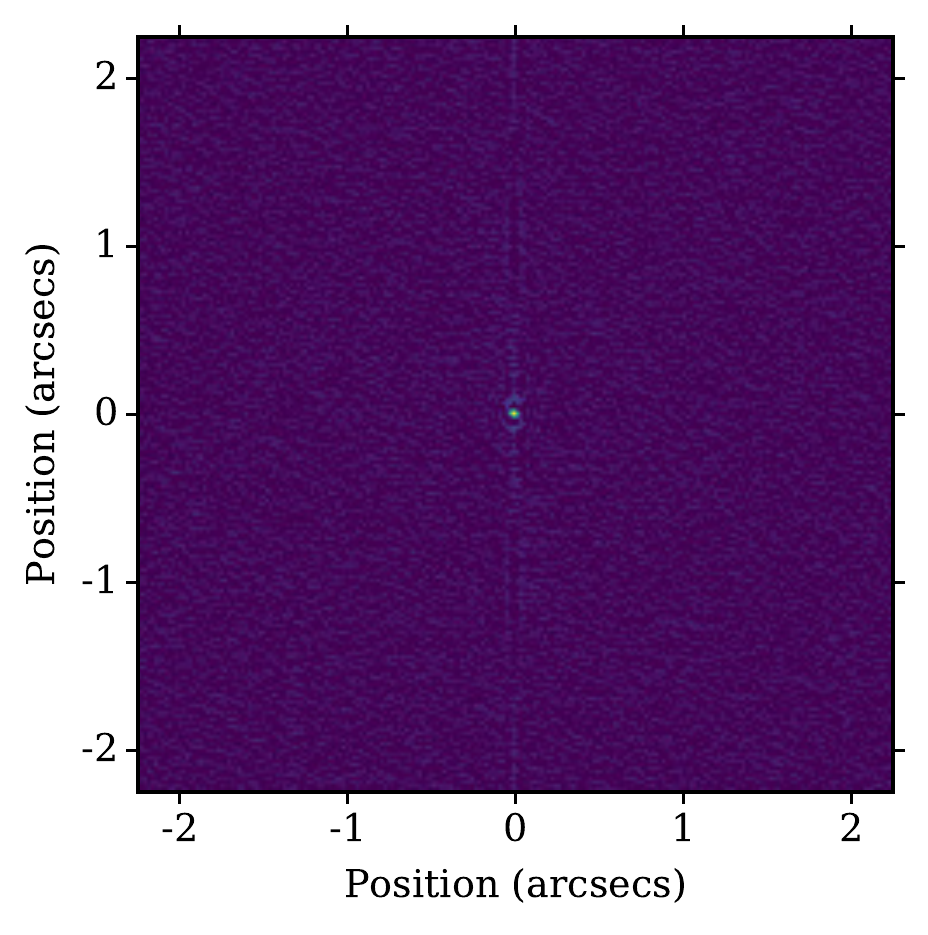}
    \caption{Blue band (562nm) NESSI speckle reconstruction showing the target star in the center. Image intensities have been square root stretched for visualization purposes.}
    \label{fig:speckleimage}
\end{figure}

\begin{figure}
	\centering
    \epsscale{1} 
    \plotone{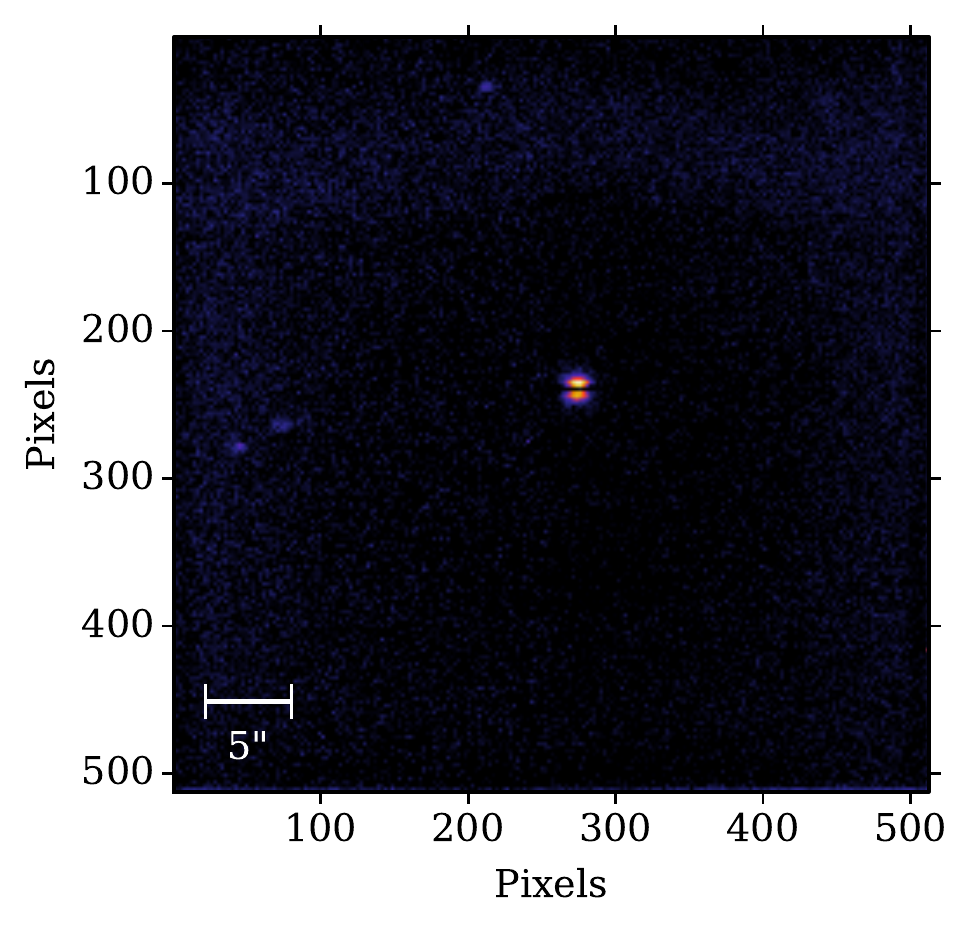}
    \caption{Keck HIRES guider frame centered on 2MASS~J06101557+2436535 showing slit centered on the star. Image intensities have been stretched for visualization purposes.}
    \label{fig:keckguider}
\end{figure}

\section{Transit validation}\label{validation}

Following the identification of a potential transiting companion around 2MASS~J06101557+2436535, we carried out additional analysis to confirm its substellar nature. We first checked light curves of nearby stars to confirm that the the source of the transits was indeed 2MASS~J06101557+2436535 and found that no nearby stars presented any significant variability. \tbf{We also checked if there was a correlation between the x and y positions of the star on the sensor and the transit event using the {\em lightkurve} package \citep{lightkurve}. We found no such correlation, suggesting that the probability of a background binary causing the transit event is low}. \tbf{We used both speckle interferometry (see Figure~\ref{fig:speckleimage}, Section~\ref{speckledata}) at close angular separations and a Keck guider image for wider separations (see Figure~\ref{fig:keckguider}) to rule out several regions of parameter space, in terms of contrast (difference in Kepler magnitudes) as a function of angular separation, for a background eclipsing binary.} Images from the Keck guider camera did not show any stellar companions with separations down to ~1.2\arcsec\ (see yellow region of Figure~\ref{fig:contrast2}). 

Based on the {\em K2}  photometry, it was possible to constrain the possibility of a background eclipsing binary (EB) based on the depth of the transit observed. As described in \citet{everett2015}, assuming a maximum eclipse depth of 50\%, the greatest difference in magnitudes that could produce the observed transit depth $\delta$ is $\Delta Kp_{max}$ = $-2.5 \log_{10}(2\delta)$. This constraint is shown by the gray region of Figure \ref{fig:contrast2} and is approximately $\Delta Kp \geq 4.66$. 

We also considered the possibility that 2MASS~J06101557+2436535 is eclipsed by a very low mass stellar companion and the light of those eclipses is diluted by a faint background star. This is statistically highly unlikely, as shown by our validation tests with \texttt{vespa} (See Section \ref{vespa}). 

To rule out false positive scenarios involving stellar companions at close separation, we obtained follow-up high resolution observations and performed additional statistical analysis to confirm its substellar nature (Sections~\ref{speckledata} and \ref{vespa}). 

\subsection{Speckle interferometry}\label{speckledata}
We carried out speckle interferometry on 2MASS J06101557+2436535 to constrain the possibility of a background eclipsing binary system causing a false positive or dilution of the transit depth. On the night of 8 November 2017 (UT), we observed our target with the NASA Exoplanet Star and Speckle Imager (NESSI), as part of an approved NOAO observing program (P.I. Howell). NESSI is a new speckle imager built for the 3.5m WIYN Telescope \citep{howell2011}, \citep{scott2017}. NESSI uses electron-multiplying CCDs (EMCCDs) to image sequences of very short (40~ms) exposures simultaneously in two bands: a “blue” band centered at 562~nm with a width of 44~nm, and a “red” band centered at 832~nm with a width of 40~nm. The fast exposures essentially freeze the atmospheric blurring and upon Fourier reconstruction, images are produced providing diffraction limited resolution.

\begin{figure}
    \epsscale{1.2}
    \plotone{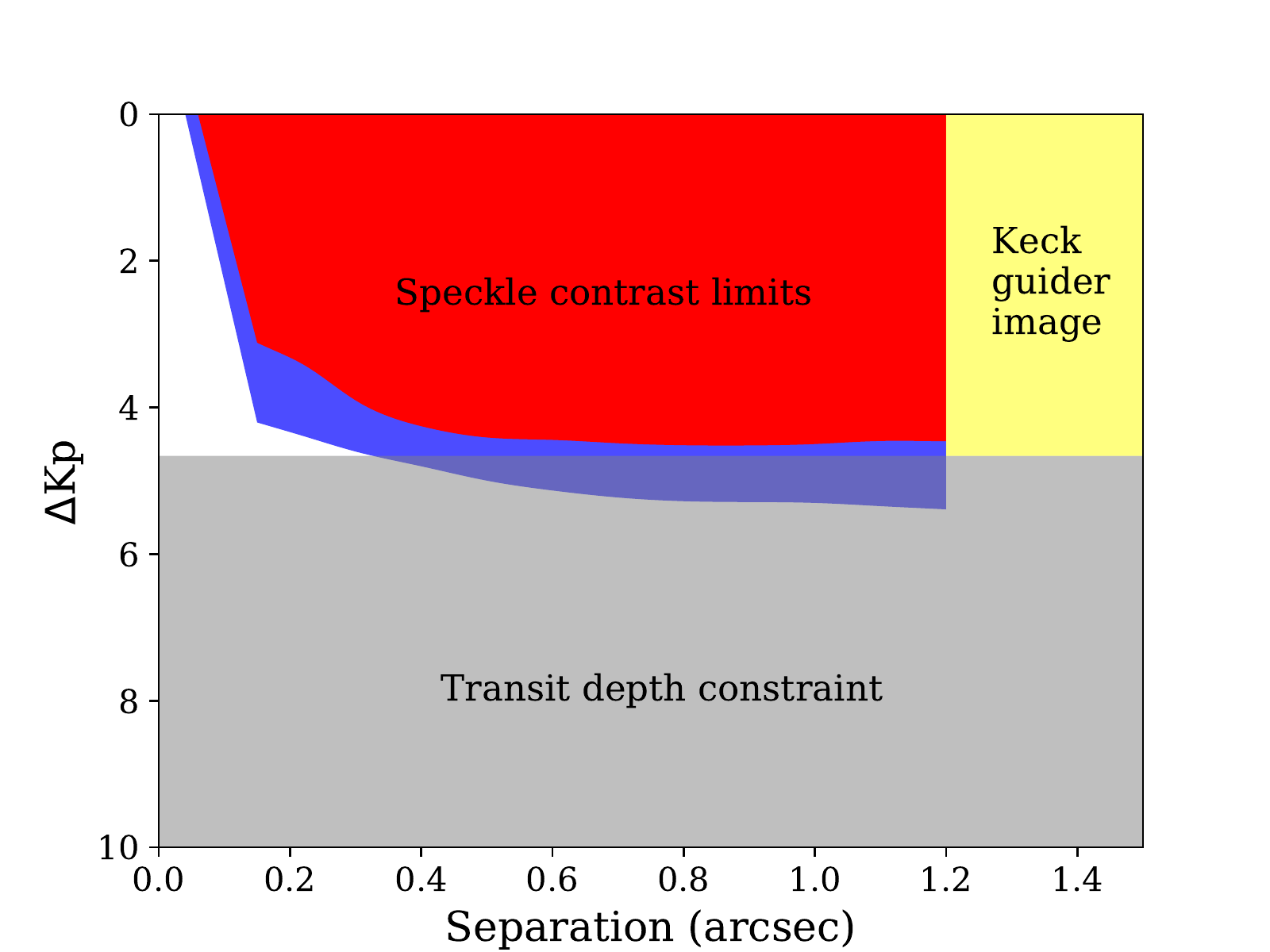}
    \caption{Contrast limits for the detection of a potential background or double star, in terms of $\Delta K_{\rm p}$, the difference in {\em Kepler} magnitude, as a function of angular separation. Colored areas are zones of exclusion. The red region represents the region of parameter space ruled out by our 832~nm speckle observations. The blue region shows additional space ruled out by the 562~nm blue speckle observations. The gray rectangle depicts the region of parameter space below which a false positive eclipsing binary would not produce a transit of the depth observed by {\em K2}. The only region of parameter space that we cannot rule out observationally is the white section to the upper left; we approach this statistically in \ref{vespa}}
    \label{fig:contrast2}
\end{figure}

For 2MASS J06101557+2436535, our imaging revealed that no secondary sources were detected; that is, we could exclude companions brighter than 1.6\% and 0.7\% of the target star, respectively, in the red and blue band at separations from 1.2\arcsec\ to as close as 30-60~mas. Figure~\ref{fig:contrast2} shows the measured background sensitivity (in a series of concentric annuli centered on the star) and the resulting 5 sigma sensitivity limit as a function of angular separation, for the two different bands. \tbf{The speckle contrast limits are in the 832~nm (red) and 562~nm (blue) bands respectively; the Kepler band contrast would be somewhere between these two.} 

\subsection{Statistical validation} \label{vespa}

We used \texttt{vespa} \citep{morton12} to further validate the planetary nature of this system. \texttt{vespa} calculates the false positive probability of a system based on several inputs, including stellar parameters, color photometry, transit photometry, and speckle/AO contrast curves. Using the aperture size used for the {\em K2} photometry as well as the contrast curves, \texttt{vespa} calculates the likelihood of each of several eclipsing binary contamination scenarios using a TRILEGAL simulation \citep{trilegal2005}, \citep{trilegal2012}. The probability of six total false positive scenarios are evaluated by \texttt{vespa}. These include background eclipsing binaries, eclipsing binary companions, hierarchical eclipsing binaries, and each of the aformentioned scenarios at double the transit period. 

We used color photometry and stellar parameters as presented in Table 1, as well as period and $R_{p}/R_{*}$ values as obtained by EXOFAST (see Section \ref{discovery}). We also input the 832~nm red band speckle contrast curve and the light curve from {\em K2}. \tbf{The \texttt{vespa}-reported false positive probability is $<10^{-4}$.} We thereby rule out contamination from a hypothetical star in the white zone of Figure \ref{fig:contrast2} resulting in the observed transits, validating the companion as either a planet or a brown dwarf.

\begin{deluxetable}{ll}
\tablecolumns{2}
\tablewidth{0pt}
\tablecaption{2MASS J06101557+2436535b Properties}
\tablehead{
\colhead{Parameter} & \colhead{Value}
}
\startdata
Period (days) & $ 7.555907^{+0.000029}_{-0.000030}$ \\
$T_0$ (BJD) & $2456784.9266^{+0.0016}_{-0.0023}$ \\
$R_{p}/R_{*}$ & $0.08270^{+0.00041}_{-0.00046}$ \\
$a/R_{*}$ & $9.7^{+1.5}_{-1.4}$ \\
Semi-Major Axis (AU) & $0.0956^{+0.0039}_{-0.0056}$ \\
Transit Depth ($\%$) & $0.6839^{+0.0067}_{-0.0075}$ \\
Transit Duration & $0.1918\pm{0.001}$ \\
Inclination $(i)$ & $85.5^{1.0}_{1.9}$ \\
Radius $(R_J)$ & $1.5^{+0.3}_{-0.3}$ \\
Eccentricity $(e)$ & $0.22^{+0.20}_{-0.15}$ \\
Linear Limb Darkening Coeff. $(u1)$ & $0.22\pm{0.05}$ \\
Quadratic Limb Darkening Coeff. $(u2)$ & $0.28\pm{0.05}$ \\
\enddata
\tablecomments{\label{tab:planetparameters} Derived planetary parameters from the transit fit.}
\end{deluxetable}

\section{Summary and discussion}\label{discussion}

During Campaign 0 of the {\em K2} mission the open cluster M35 was targeted using a superstamp region encompassing the bulk of the cluster. Performing a search for transits in this M35 superstamp, we have found a previously unappreciated A-type star with a transiting companion that is very likely substellar. The host star, 2MASS J06101557+2436535, is a rapidly rotating star with a spectral type A2-A3 based on high resolution spectra taken with Keck/HIRES. 

We conducted ground-based transit follow-up and statistical analysis to rule out most false positive scenarios. Speckle imaging with the 3.5m WIYN telescope and NESSI revealed no stellar sources other than 2MASS~J06101557+2436535 within the contrast limit. A partial transit observed with the 0.4m LCO telescope on Haleakala corroborated the data from {\em K2}, showing a transit at the expected depth and time. We have also presented an analysis with the code \texttt{vespa}, \citep{morton12}, which yields a false positive probability of $<10^{-4}$, validating the system as a substellar companion to a hot star. The derived planetary radius is 1.5~$R_{\rm J}$.

This system is unique for a number of reasons, particularly if the companion object turns out to be planetary mass. It is one of a very small number of transiting objects orbiting A-type hosts. The temperature range of 2MASS~J06101557+2436535 also makes it one of the hottest stars with a substellar companion \citep{gaudi2017}. Among transiting planet hosts, only KELT-9 \citep[$\sim$10200~K;][]{joner2017} and possibly KELT-20/MASCARA-2 \citep[$\sim$8700~K][]{lund2017} are hotter. The directly imaged planet host kappa Andromedae is also hotter, but the mass of the companion is uncertain. There are also several subdwarf B stars which have been claimed to have planets based on pulsation timing, though these are also controversial. \tbf{We also note that the host star has the highest projected rotational velocity  of any host star with a transiting substellar companion at $v$ sin $i$ = 193.7 $km\ s^{-1}$, the second being KELT-21 \citep{keltjohnson2018} at 146 $km\ s^{-1}$. This poses a large challenge to spectral characterization of the host, which we have described in Section~\ref{spectraltype}.}

Mid-infrared secondary eclipse studies may provide additional confirmation of the substellar nature of 2MASS J06101557+2436535b. Its estimated equilibrium temperature of $\sim$2000~K implies an eclipse depth of $\sim$1500~ppm, well within reach of the {\em Spitzer} Space Telescope, provided that the eccentricity can be constrained. Because the host is an A type star, it is very rapidly rotating and a prime candidate for Doppler tomographic follow up observations. While we have statistically validated this system, Doppler tomography would further exclude a background eclipsing binary system. In addition, Doppler tomography provides a measurement of spin-orbit angle. Nearly every hot Jupiter around an A or earlier type star has been found to be spin-orbit misaligned (\citet{talens2017}, \citet{zhou2016}, and \citet{collier2010}, although this may be biased by a preference for targets with lower $v\sin i$ values. A measurement of spin-orbit angle for this system will add to the growing number of such measurements that can shed light on planet formation and migration history.

\section{Acknowledgements}\label{acknowledgements}

We thank Nic Scott for assistance with speckle observations. John Stauffer, Lynne Hillenbrand, Trevor David, Geert Barentsen, and Christina Hedges provided helpful discussion.
This paper makes use of EXOFAST (Eastman et al.~2013) as provided by the NASA Exoplanet Archive, which is operated by the California Institute of Technology, under contract with the National Aeronautics and Space Administration under the Exoplanet Exploration Program. This work also made use of PyKE (Still \& Barclay 2012), a software package for the reduction and analysis of Kepler data. This open source software project is developed and distributed by the NASA Kepler Guest Observer Office. 
This work has made use of data from the European Space Agency (ESA) mission
{\em Gaia} (\url{https://www.cosmos.esa.int/gaia}), processed by the {\em Gaia}
Data Processing and Analysis Consortium (DPAC,
\url{https://www.cosmos.esa.int/web/gaia/dpac/consortium}). Funding for the DPAC has been provided by national institutions, in particular the institutions
participating in the {\em Gaia} Multilateral Agreement.
This work makes use of observations from the LCOGT network. 
Finally, the authors wish to recognize and acknowledge the very significant cultural role and reverence that the summit of Maunakea has always had within the indigenous Hawaiian community. We are most fortunate to have the opportunity to conduct observations from this mountain.

\facilities{WIYN, Keck:I (HIRES), LCOGT}
\software{DAOphot \citep{stetson87}, IRAF \citep{tody1986},\citep{tody1993}, PyKE \citep{still12}, EXOFAST \citep{eastman2013}, AstroImageJ \citep{astroimagej}, emcee \citep{emcee2013}, vespa \citep{morton12}, Astropy \citep{astropy2013}, Matplotlib \citep{hunter2007}}

\newpage

\bibliography{references}
\end{document}